\documentclass[12pt]{article}
\pdfoutput=1 
 
\usepackage{hyperref}    
\usepackage{amsmath}      
\usepackage{amssymb} 
\usepackage{graphicx}
\usepackage{url}

\usepackage{float,wrapfig,color} 

\allowdisplaybreaks
\def\eq#1{(\ref{#1})}
\def\s[#1\s]{\begin{align}\begin{split}#1\end{split}\end{align}}
\def\[#1\]{\begin{align}#1\end{align}}

\begin{document}

\begin{titlepage} 

\title{
\hfill\parbox{4cm}{ \normalsize YITP-21-36}\\ 
\vspace{1cm} 
Phase profile of the wave function of canonical tensor model
and emergence of large spacetimes 
}

\author{
Naoki Sasakura\footnote{sasakura@yukawa.kyoto-u.ac.jp}
\\
{\small{\it Yukawa Institute for Theoretical Physics, Kyoto University,}}
\\ {\small{\it  Kitashirakawa, Sakyo-ku, Kyoto 606-8502, Japan}}
}


\maketitle

\begin{abstract}
To understand spacetime dynamics in the canonical tensor model of quantum gravity
for the positive cosmological constant case,
we analytically and numerically study the phase profile of its exact wave function 
in a coordinate representation, instead of the momentum representation analyzed so far.
A saddle point analysis shows that Lie group symmetric spacetimes are strongly favored due to 
abundance of continuously existing saddle points, giving an emergent fluid picture. 
The phase profile suggests that spatial sizes grow in ``time", 
where sizes are measured by the tensor-geometry correspondence previously introduced 
using tensor rank decomposition. 
Monte Carlo simulations are also performed for a few small $N$ cases 
by applying a re-weighting procedure to an oscillatory integral which expresses the wave function.
The results agree well with the saddle point analysis, 
but the phase profile is subject to disturbances in a large spacetime region, 
suggesting existence of light modes there and 
motivating future computations of primordial fluctuations from the perspective of 
canonical tensor model.
\end{abstract}

\end{titlepage}

\section{Introduction}
\label{sec:introduction}
An open question in fundamental theoretical physics is how to quantize gravity. 
In addition to sophisticated applications of quantum field theoretical methods to general 
relativity \cite{Reuter:2019byg,Eichhorn:2018yfc},
various discretized approaches have been pursued 
\cite{Loll:2019rdj,Rovelli:2014ssa,Surya:2019ndm,Konopka:2006hu,
Wolfram:2020jjc,Trugenberger:2016viw,Akara-pipattana:2021zzy}.
A challenge in such discretized approaches is whether continuum spacetime-like
entities can be generated and general relativity is obtained as effective theory. 
One of such discretized approaches is 
the tensor model \cite{Ambjorn:1990ge,Sasakura:1990fs,Godfrey:1990dt,Gurau:2009tw}, 
which was proposed hoping to extend the success of 
the matrix model \cite{DiFrancesco:1993cyw} 
for two-dimensional quantum gravity to higher dimensions.
However, the tensor model suffers from dominance of singular 
spaces (dominance of branched polymers \cite{Bonzom:2011zz,Gurau:2011xp}), 
which cannot be regarded as macroscopic continuous spacetimes like the universe.

An interesting direction of study to overcome the issue is given by incorporating a temporal direction.
Causal dynamical triangulation has been shown to generate macroscopic 
spacetime-like entities \cite{Ambjorn:2004qm}, while this is hard for dynamical triangulation. 
The main difference between the two models is that the former possesses a temporal direction,
while the latter does not. 
This success motivated the current author to construct a new tensor model 
in Hamiltonian formalism, 
which we call canonical tensor model (CTM) \cite{Sasakura:2011sq,Sasakura:2012fb}. 
This is a constrained system with a number of first-class constraints, which are
analogs to the Hamiltonian and momentum constraints in the ADM formalism of general
relativity \cite{Arnowitt:1962hi}.

CTM has been shown to have various interesting properties classically and quantum mechanically.
Classically, there exist a few connections between CTM and general relativity. 
One is that the $N=1$ case agrees with the mini-superspace approximation 
of general relativity \cite{Sasakura:2014gia}. 
Another is that, in a formal continuum limit with $N=\infty$, 
the classical dynamics of CTM can be 
described by a Hamilton-Jacobi equation in general relativity coupled with a scalar and higher spin fields
\cite{Chen:2016ate}. 
As for quantization, it is straightforward to canonically quantize CTM \cite{Sasakura:2013wza}, 
and an exact physical state wave function in a general form has been obtained \cite{Narain:2014cya}. 
An eminent property of the wave function
is that the wave function has peaks (or ridges) at the locations where the tensor argument takes values
symmetric under Lie groups \cite{Obster:2017dhx,Obster:2017pdq}. 
This property, which may be called Lie group symmetry emergence by quantum coherence, 
would be a key to 
spacetime emergence, since the actual spacetime can be featured by 
various Lie group symmetries like Lorentz, de Sitter, 
gauge, and so on.  There are some other aspects of CTM, which are summarized for instance in 
an appendix of \cite{Sasakura:2019hql}.

However, in the results listed above, there exists an inconsistency concerning which of $Q_{abc}$ or $P_{abc}$, namely, the canonical conjugate variables of CTM, represents spacetimes. 
More precisely, in the $N=1$ agreement mentioned above,  $Q_{111}$ represents the spatial volume 
in general relativity,
while, in the formal continuum limit mentioned above, $P_{abc}$ is associated to the spacetime metric. 
In the latter case, the main reason why we obtain the Hamilton-Jacobi equation, which is a first-order 
differential equation in time, rather than the genuine second-order one in general relativity, 
comes from the fact that the Hamiltonian constraint of CTM is at most linear in $Q_{abc}$.  
Rather, $Q_{abc}$ would be more entitled to be associated with spacetimes
to correctly produce the second order differential equation of general relativity,
since the Hamiltonian constraint of CTM is quadratic in $P_{abc}$. 
 
The main motivation of this paper is to straighten out the above inconsistent treatment of CTM so far.
We consistently regard $Q_{abc}$ as representing spacetimes to understand the spacetime 
dynamics of CTM.
Studying the classical equation of motion
of $Q_{abc}$ in CTM, however, does not seem to lead to any physically sensible outcomes,
since the classical dynamics would be able to start and end with arbitrary values of $Q_{abc}$.
Rather, we employ quantized CTM and study the wave function in $Q_{abc}$ representation
to restrict ourselves to high-probabilistic configurations at the wave function peaks.
The Lie group symmetry emergence mentioned above, which was argued in the $P_{abc}$
representation, will provide similar selections of high-probabilistic values of $Q_{abc}$.
As for the cosmological constant parameter in CTM \cite{Sasakura:2014gia}, 
we exclusively consider the positive case, since 
the previous analyses of the negative case \cite{Sasakura:2019hql,Lionni:2019rty} and 
of a toy matrix model \cite{Sasakura:2020rqz} clearly indicate that the positive case has 
the only possibility of sensible emergent spacetimes.\footnote{More precisely,
it is argued that emergent spaces can have distinct integer dimensions and symmetries
only for the positive cosmological constant case \cite{Sasakura:2020rqz}.}
 
In Section~\ref{sec:setup}, we first define the setup, namely, the wave function in the $Q_{abc}$ representation,
which has the form of an oscillatory multiple integration.
The wave function is treated by a saddle point method.
Interestingly, the saddle point equation is found to be a coupled system of a tensor 
eigenvalue/vector equation \cite{Qi} and a tensor rank decomposition \cite{Landsberg2012}.
In Section~\ref{sec:lie}, we argue that, when $Q_{abc}$ takes a value symmetric under a Lie group,
there appears abundance of continuously existing saddle points along a group orbit of the Lie group. 
These continuously existing saddle points coherently enhance the value of the wave function, 
so that the wave function has peaks (or ridges) along Lie group symmetric values of $Q_{abc}$. 
We also argue that, for such a value of $Q_{abc}$,
the system can effectively be described in a continuous manner with 
an emergent fluid picture in which microscopic variables are averaged.
In Section~\ref{sec:profile} and \ref{sec:trd}, 
we discuss the phase profile of the wave function based on the saddle point analysis,
and give a qualitative argument that sizes of spaces grow in ``time".
Here spatial sizes are measured by the tensor-geometry correspondence 
that was previously introduced in \cite{Kawano:2018pip} using the tensor rank decomposition.
In Section~\ref{sec:montecarlo}, we show the results of Monte Carlo simulations which were obtained
by applying the re-weighting procedure to a few small $N$ cases.  
The results are consistent with the saddle point analysis. We also observe that the phase profile
is subject to disturbances in a large spacetime region, suggesting presence of light modes there. 
The last section is devoted to a summary and an outlook.     
 
\section{The wave function in $Q$ representation and its saddle points}
\label{sec:setup}
Dynamical variables of the canonical tensor model (CTM)
are given by a canonical conjugate pair of real symmetric
three-indices tensors, $Q_{abc}$ and $P_{abc}$\ ($a,b,c=1,2,\ldots,N$)
\cite{Sasakura:2011sq,Sasakura:2012fb}. 
The quantization of CTM is straightforward by the standard canonical quantization 
with $[\hat Q_{abc},\hat P_{def}]=i\, \delta_{abc,def}$ \cite{Sasakura:2013wza}.
The  physical states of quantized CTM are defined by the following physical state condition,
\[
\hat {\cal H}_a |\Psi \rangle =\hat {\cal L}_{ab} |\Psi\rangle=0,
\label{eq:physcond}
\]
where $\hat {\cal H}_a$ and $\hat {\cal L}_{ab}$ are the quantized first-class constraints of CTM.
These constraints are respectively the analogs to the Hamiltonian and the momentum constraints of the 
ADM formalism of general relativity \cite{Arnowitt:1962hi}, 
and in particular the first one is the analogue to the Wheeler-DeWitt equation \cite{DeWitt}.
An exact solution in a general form to the physical state condition is known and the wave function in the $P$ representation is given by \cite{Narain:2014cya}
\s[
\Psi(P)&:=\langle P | \Psi \rangle =\varphi(P)^R,
\label{eq:defofPsi}
\s]
where $R=(N+2)(N+3)/4$, and\footnote{We need some cares to rigorously define this oscillatory integral. 
The most rigorous way would be to take Lefschetz thimbles as integration contours \cite{Witten:2010cx}.
Or we may also introduce a regularization parameter and take its zero limit \cite{Obster:2017dhx}.} 
\[
\varphi(P):= \int_{\cal C} d\tilde \phi \int_{\mathbb{R}^{N}} \prod_{a=1}^N d\phi_a 
\, e^ 
{
I \left(
P_{abc}\phi_a \phi_b \phi_c  - k \, \phi_a \phi_a \tilde \phi + \tilde \phi^3
\right)
},
\label{eq:defofvarphi}
\]
where $I$ denotes the imaginary unit, $I^2=-1$, and pairwise lower indices are always assumed to be 
summed over, unless otherwise stated.  
The parameter $k$ has a relation $k^3 \propto \lambda$, where $\lambda$ 
is a parameter of CTM corresponding to the cosmological constant of general 
relativity \cite{Sasakura:2014gia}. 
Throughout this paper, $k$ is assumed to be positive, corresponding to the positive cosmological constant case. 
The reason for the restriction is that the Lie group symmetry emergent phenomenon
mentioned in Section~\ref{sec:introduction} 
is evident only for the positive case \cite{Obster:2017dhx,Obster:2017pdq}.
In fact, the analysis of a simplified toy wave function performed in \cite{Sasakura:2020rqz}
shows that emergence of spaces can only occur in sensible manners 
for the positive cosmological constant 
case.\footnote{See a footnote in Section~\ref{sec:introduction}.} 

\begin{figure}
\begin{center}
\includegraphics[width=4cm]{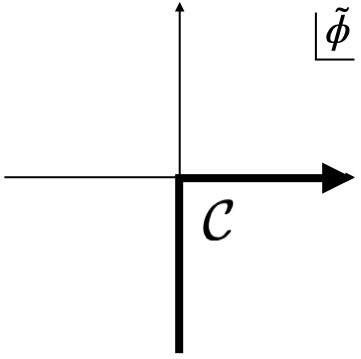}
\caption{The integration contour ${\cal C}$ of $\tilde \phi$.}
\label{fig:intcont}
\end{center}
\end{figure}

The integration contour ${\cal C}$ of $\tilde \phi$ in \eq{eq:defofvarphi} is taken as in
Figure~\ref{fig:intcont}. 
The integration contour in the previous works is taken on the real axis as the primary simplest choice, 
with which the integration over $\tilde \phi$ leads to the Airy Ai function, 
${\rm Ai}(-x)$ with $x=3^{-1/3} k\, \phi^2$.
This is a real oscillatory function and hence the wave function represents a standing wave, which 
is supposed to represent a superposition of shrinking and expanding universes from 
physical viewpoints.
In this paper, however, we take a different choice given in Figure~\ref{fig:intcont}, which is 
supposed to represent an expanding 
universe.\footnote{Since we do not have an explicit time variable, which of expanding or shrinking
would be a matter of convention.}
With this choice, the integration over $\tilde \phi$ leads to ${\rm Ai}(-x)+I\, {\rm Bi}(-x)$. 
which has the asymptotic behavior of an advancing wave, $\sim e^{-(2/3) I x^{3/2}}$.

Now let us write down the wave function in the $Q$ representation
from the expressions \eq{eq:defofPsi} and \eq{eq:defofvarphi}.\footnote{We use 
the same symbol $\Psi$ for both $Q$ and $P$ representations, discriminating them by the argument.}
We obtain
\s[ 
\Psi(Q)&:=
\int_{{\mathbb R}^{\# P}} \prod_{a\leq b \leq c=1}^N dP_{abc}\ e^{-I P_{abc}Q_{abc}}\, \Psi(P) \\
&=\int_{{\mathbb R}^{\# P}}  \prod_{a\leq b \leq c=1}^N dP_{abc} 
\prod_{j=1}^R 
\int_{{\mathbb R}^N}
\prod_{a=1}^N d\phi_a^j
\int_{\cal C}
d\tilde \phi^j  \,
e^{ I\left( - P_{abc}Q_{abc}+ \sum_{i=1}^R \left(
P_{abc}\phi^i_a \phi^i_b \phi^i_c  - k \, \phi^i_a \phi^i_a \tilde \phi^i + (\tilde \phi^i)^3\right)
\right)
},
\label{eq:psiq}
\s]
where we have assumed $R$ to be integer and have rewritten the power $R$ in 
\eq{eq:defofPsi} by introducing $R$ replicas of $\phi_a,\tilde \phi$.  
This assumption of integer $R$ 
restricts the possible values of $N$ but would not be essential for large $N$ cases.  
In the Monte Carlo simulations in Section~\ref{sec:montecarlo}, we take $R$ to be one of the 
nearest values to the real values for $N=3$ and $4$.
In \eq{eq:psiq}, we could have integrated over $P_{abc}$ to get an expression with delta functions,
but we have not done this for the sake of the discussions below.
$\# P$ denotes the number of the independent
components of $P_{abc}$, i.e. $N(N+1)(N+2)/6$.
Note that summation over replica indices $i$ must be explicitly indicated,
whenever summation is needed.

The saddle point equation for \eq{eq:psiq} is given by the following simultaneous equations,
\s[
&P_{abc} \phi^i_b \phi^i_c=\frac{2 k \tilde \phi^i}{3} \phi^i_a, \\
&Q_{abc}=\sum_{i=1}^R \phi_a^i\phi_b^i\phi_c^i, \\
&k \phi_a^i \phi_a^i=3 (\tilde \phi^i)^2,
\label{eq:saddleeq}
\s]
where $Q_{abc}$ is the only external variable.  
The equations are obtained by taking partial derivatives of the exponent in \eq{eq:psiq} with respect to $\phi_a^i$, $P_{abc}$ and $\tilde \phi^i$, respectively.
Interestingly, the first equation is the tensor eigenvalue/vector equation \cite{Qi}
and the second has the form of the tensor rank decomposition \cite{Landsberg2012}, 
respectively. These equations are recently attracting some attentions 
in the field of applied mathematics due to their applications to data analysis 
(See for instance the introduction of \cite{Evnin:2020ddw} for a concise account.). 
As can be seen in the third equation of \eq{eq:saddleeq}, 
the positivity of $k$, corresponding to the positive 
cosmological constant \cite{Sasakura:2014gia}, is necessary for the existence of real 
solutions.\footnote{We assume that non-real saddle points are subdominant and can hence be ignored. 
In fact, for negative $k$, the Lie group emergent phenomenon is much less evident than 
the positive case \cite{Obster:2017dhx,Obster:2017pdq}.}
Note that, because of the contour ${\cal C}$ shown in Figure~\ref{fig:intcont},
the solution for $\tilde \phi^i$ to the third equation should be taken in the positive branch,
namely, $\tilde \phi^i\geq0$.
By putting \eq{eq:saddleeq} to the exponent of \eq{eq:psiq}, 
the saddle point approximation to the wave function in the lowest order is given by
\s[
\Psi(Q)\sim \sum_{\rm Solutions} e^{-2 I \sum_{i=1}^R (\tilde \phi^i)^3}.
\label{eq:psisaddle}
\s]

Let us estimate $\Psi(Q)$ for general values of $Q_{abc}$, 
where ``general" means that $Q_{abc}$ is taking a value with no particular structure. 
In \eq{eq:saddleeq}, there are totally $NR+\# Q+R$ conditions, while the total number of variables is
$\# P+NR+R$. Since $\# Q=\# P$, the numbers of the conditions and the variables are equal,
 and only a finite discrete set of solutions to \eq{eq:saddleeq} will exist in general. 
 Then, \eq{eq:psisaddle} implies 
 \s[
 \Psi(Q)\sim O(1) \hbox{ for general }Q_{abc}.
 \label{eq:estimate}
 \s]
 
\section{The wave function at Lie group symmetric $Q_{abc}$}
\label{sec:lie}
The estimation \eq{eq:estimate} may change, if $Q_{abc}$ takes a value which allows a macroscopic 
number\footnote{Compared to $N$ or so.} or even a continuous set of 
saddle point solutions. However, configurational abundance of this sort is not enough 
for \eq{eq:psisaddle} to become large in general, 
because it is also necessary that the sum over the solutions in \eq{eq:psisaddle}
coherently contribute, rather than cancel among them due to oscillatory phases.

Such coherence would only occur, when $Q_{abc}$ has particular structure.  
Let us consider $Q_{abc}=Q^0_{abc}$ which is invariant under a Lie group $G$,
namely,   $Q^0_{abc}=r(g)_a^{a'} r(g)_b^{b'} r(g)_c^{c'} Q^0_{a'b'c'}$ for an orthogonal representation 
$r(g)$ of a Lie group $G$.
For a given such $Q^0_{abc}$, let us suppose that there is a solution to \eq{eq:saddleeq}, 
${\phi^0}_{a}{}^i$, ${\tilde \phi^0}{}^i$, and $P_{abc}^0$. 
Then, this can be extended to a continuous set of solutions along a group orbit of $G$ by
$r(g)_a^{a'}{\phi^0}_{a'}{}^i$ and $r(g)_a^{a'} r(g)_b^{b'} r(g)_c^{c'} P^0_{a'b'c'}$,  
because of the invariance of $Q^0_{abc}$ under $G$.
An important point is that ${\tilde \phi^0}{}^i$ takes the same value along the group orbit, 
and therefore the sum over the solutions is coherent in \eq{eq:psisaddle}. We obtain  
\s[
\Psi_{\rm naive}(Q^0)\sim V_{\rm orbit} \, e^{-2 I \sum_{i=1}^R  ({\tilde \phi^0}{}^i)^3},
\label{eq:naive}
\s] 
where $V_{\rm orbit}$ denotes the volume of the group orbit. 

However, the enhancement in the preceding paragraph is naive. 
As we will see below, we can get much stronger enhancement, 
if we consider a solution which has $P^0_{abc}$ also invariant under $G$. 
Note that this requirement is in harmony with the Lie group symmetry emergent phenomenon 
of the wave function in the $P_{abc}$ representation, which is mentioned in
Section~\ref{sec:introduction}.
 
To estimate $\Psi(Q)$ under the additional requirement of the invariance of $P^0_{abc}$ under $G$
for saddle point solutions, let us first perform the rescaling $\phi_a^i=\tilde \phi^i v_a^i$ for convenience.  
Ignoring the solutions with $\tilde \phi^i=0$ as being 
irrelevant\footnote{From \eq{eq:saddleeq}, we also get $\phi_a^i=0$ in this case.}, 
the saddle point equation 
\eq{eq:saddleeq} can be rewritten as
\s[
&P_{abc} v^i_b v^i_c=\frac{2 k}{3} v^i_a, \\
&Q_{abc}=\sum_{i=1}^R (\tilde \phi^i)^3 v_a^i v_b^i v_c^i, \\
&v_a^i v_a^i=\frac{3}{k}.
\label{eq:saddlee}
\s]
Puitting $P_{abc}=P^0_{abc}$ invariant under $G$, 
the first and the third equations of \eq{eq:saddlee} imply
that the solutions for $v_a^i$ exist along the group orbit,
\[
v_a(g):=r(g)_a^{a'} v_{a'}^0,
\]
where $v_{a}^0$ is a representative solution with $v_a^0v_a^0=3/k$. 
By multiplying $v^0_a$ to the first equation of \eq{eq:saddlee}, we see that the size 
of $P_{abc}^0$ is implicitly determined by $v_a^0$ as 
\[
P^0_{abc}  v_{a}^0 v_{b}^0v_{c}^0=2.
\] 
Finally, the second equation of \eq{eq:saddlee} requires
\[
Q_{abc}^0=\sum_{i=1}^R (\tilde \phi^i)^3 v_a(g_i)  v_b(g_i)  v_c(g_i). 
\label{eq:qeqg}
\] 

Let us summarize the general procedure to find saddle point solutions above. 
We first parameterize Lie group invariant $P^0_{abc}$ and obtain an eigenvector $v^0_a$
consistently with the normalization conditions.
Then, tune $P^0_{abc}$ so that \eq{eq:qeqg} is satisfied with a choice of $g_i$'s and $\tilde \phi^i$'s.
Here note that we may generally have more abundance of solutions than 
what leads to \eq{eq:naive}, because we have the freedom to simultaneously change $g_i$'s
on the group orbit and the values of $\tilde \phi^i$'s, as long as \eq{eq:qeqg} is satisfied.
 
To understand the abundance more clearly, 
let us describe \eq{eq:qeqg} in a continuous manner like a fluid.
Namely, we assume $R$ is large enough to allow us to express the locations of $g_i$ collectively
in terms of a smooth non-negative function $\rho(g)$:\footnote{The original expression can be 
recovered by putting $\rho(g)= \frac{1}{R}\sum_{i=1}^R \delta(g-g_i)$ 
to \eq{eq:qeqgcont}.}
\s[
\sum_{i=1}^R \rightarrow R \int_G dg\, \rho(g),
\s]
where $dg$ denotes the Haar measure of $G$ with the normalization, $\int_G dg \rho(g)=1$. 
 We also assume that $\tilde \phi^i$ can be described by a smooth function 
$\tilde \phi(g)\geq 0$. 
Note that the reason for $\tilde \phi(g)\geq 0$ comes from $\tilde \phi^i\geq 0$, 
that is commented above \eq{eq:psisaddle}.
Then we obtain 
 \[
Q_{abc}^0=R\int_G dg\, \rho(g)\tilde \phi(g)^3\, v_a(g)  v_b(g)  v_c(g),
\label{eq:qeqgcont}
\]
as a continuum analogue of \eq{eq:qeqg}.
In general the representation of $G$ may not be faithful on $v^0_a$, but
this can easily be taken care of by imposing $\rho(g)=\rho(g')$ and $\tilde \phi(g)=\tilde \phi(g')$ 
for $v_a(g)=v_a(g')$.

The invariance of $Q^0_{abc}$ under $G$ for the expression \eq{eq:qeqgcont} can be satisfied by 
a solution,
\[
&\rho(g)\,\tilde \phi(g)^3=\sigma
\label{eq:invsol}
\]
with a constant $\sigma\geq 0$ along ${}^\forall g\in G$, leading to
\[
Q_{abc}^0=R\sigma \int_G dg\,v_a(g)  v_b(g)  v_c(g).
\label{eq:qtov}
\]
The invariance of $Q_{abc}^0$ can be proven 
from $r(g')_{a}^{a'} v_{a'}(g)=v_{a'}(g'g)$ and the invariance of the Haar measure.
We may find the solution which satisfies \eq{eq:qtov} by tuning $P^0_{abc}$ and $\sigma$.

The most important point in the solution above is that $\rho(g)$ and $\tilde \phi(g)$ appear 
only in the combination $\rho(g)\,\tilde \phi(g)^3$. 
Since they are independent degrees of freedom\footnote{$\rho(g)$ comes from the locations of $g_i$'s, and $\tilde \phi(g)$ from $\tilde \phi^i$'s.}, 
they have the freedom to  be rescaled 
as long as $\rho(g)\,\tilde \phi(g)^3=\sigma$. 
Since this freedom of rescaling exists locally on the gauge orbit $g\in G$, 
the solutions are much more abundant than the naive one leading to \eq{eq:naive}.
In addition, what is remarkable is that the sum in \eq{eq:psisaddle} is coherent, because
the exponent  of the summand is also expressed by this combination,
\[
\sum_{i=1}^R (\tilde \phi^i)^3\rightarrow 
R \int_G dg \, \rho(g)\tilde \phi(g)^3 =R \sigma,
\label{eq:phasegroup}
\]
where we have used \eq{eq:invsol}. Thus we obtain
\[
\Psi\left(Q^0\right)\sim V \,  e^{-2 I R \sigma },
\label{eq:psigroup}
\]
where $V$ denotes the volume of the solutions, 
that should be much larger than $V_{\rm orbit}$ in \eq{eq:naive}.
It is important to explicitly compute $V$, but we leave it for future study. 

\section{Phase profile of the wave function and large spacetimes}
\label{sec:profile}
Though we do not know well the volume factor in \eq{eq:psigroup}, we rather have good 
knowledge about the phase factor at least in the lowest order approximation of the saddle point method.
In ordinary cases, one can know the direction of time evolutions from a phase profile of a wave function,
because it is transverse to constant phase surfaces.
The purpose of this section is to discuss the ``time" evolution in CTM by assuming this 
ordinary fact.
A subtlety of this assumption is that ``time" is a subtle issue in CTM as it is in Wheeler-DeWitt equation, 
and we need more study to definitely discuss the connection between the ``time" evolution 
and the phase profile of the wave function in CTM. We leave this subtle problem for future study, and 
restrict ourselves to working under the assumption.

\begin{figure}
 \begin{center}
 \includegraphics[width=7cm]{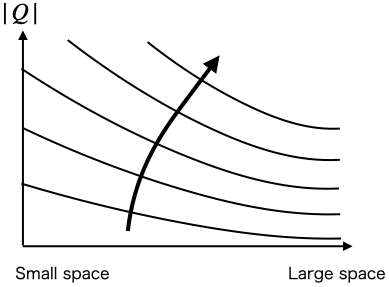}
 \caption{A typical phase profile of the wave function of CTM. Constant phase surfaces are drawn by solid lines.
 A flow line transverse to the lines is drawn, which is expected to describe a time evolution in CTM.  }
 \label{fig:wave}
 \end{center}
 \end{figure}

By using \eq{eq:qtov} and \eq{eq:psigroup}, one obtains
\[
\Psi\left(Q^0\right)\sim V
 \, \exp \left(-\frac{2 I \left|Q^0\right|}{|\int_G dg\,v_a(g)  v_b(g)  v_c(g)|}\right),
\]
where the norm of a tensor is defined by $|Q|=\sqrt{Q_{abc}Q_{abc}}$.
Since $v_a(g)$ has a constant size (see the third equation of \eq{eq:saddlee}),  
one can roughly argue that 
$|\int_G dg\,v_a(g)  v_b(g)  v_c(g)|$ tends to take a smaller value for a more extended group orbit.
This is because, as a gauge orbit is more extended,
 directions of $v_a(g)$ are more widely distributed and the integral 
$\int_G dg\,v_a(g)  v_b(g)  v_c(g)$ tends to contain more cancellations. 
Therefore a typical phase profile of the wave function of CTM can be illustrated as in Figure~\ref{fig:wave}. 
Here we identify a gauge orbit $v_a(g)\ ({}^\forall g\in G)$ as a space represented by $Q^0_{abc}$,
while more discussions about the correspondence between tensor and geometry 
will be given in Seciton~\ref{sec:trd}. 
From the relation between the $N=1$ CTM and the mini-superspace approximation of general 
relativity \cite{Sasakura:2014gia},
it would be reasonable to assume that $|Q|$ has a positive correlation with time.
Therefore one can expect that a space becomes larger in time as is shown by a flow line in 
Figure~\ref{fig:wave}.

Let us check the above general argument in concrete simple examples with $G=SO(N-1)$. 
Let us parameterize $P_{abc}$ with $G=SO(N-1)$ as follows:
\s[
&P^0_{111}=a, \\
&P^0_{1ii}=b, \ \ i=2,3,\ldots,N,
\s] 
where $a$ and $b$ are real and $b\neq 0$ for later convenience. 
Let us parameterize the eigenvector $v^0_a$ by
\[
v^0=\frac{2k}{3} (x,\vec y),
\]
where $x$ corresponds to $v^0_1$,
and $\vec y$ to the $N-1$ dimensional part $v_i^0\ (i=2,3,\ldots,N)$.  
The non-zero solution to the eigenvector equation in \eq{eq:saddlee} is given by
 \s[
 &x=\frac{1}{2b}, \\
 &|y|=\sqrt{2} |x| \sqrt{1-a x}.
 \s]
One can see that $x$ and $|y|$ can be tuned to any real values except $x=0$ 
by tuning $a$ and $b\neq 0$.

Now let us impose \eq{eq:qtov}.
From the symmetry, it is obvious that only $Q^0_{111}$ and $Q^0_{1ii}\ (i=2,3,\ldots N)$ 
take non-zero values. We obtain
\s[ 
&Q^0_{111}=R \sigma \left( \frac{2k}{3}\right)^3 x^3,\\
&Q^0_{1ii}=R \sigma \left( \frac{2k}{3}\right)^3 \frac{x |y|^2}{N-1},\ (i=2,3,\ldots,N),
\label{eq:qx}
\s]
where we have used $\int_G dg=1$ and $\sum_{i=2}^N Q^0_{1ii}=R \sigma (2k/3)^3 x|\vec y|^2$.
Note that, for $x, \vec y \neq0$, $Q^0_{111}$ and $Q^0_{1ii}$ have the same sign. 
Solving them for $x$ and $\vec y$, and putting them into $x^2+|\vec y|^2=27/4 k^3$, which is 
the normalization condition in \eq{eq:saddlee}, we obtain
\s[
R \sigma=\left( \frac{k}{3} \right)^\frac{3}{2} Q^0_{111} 
\left( 1+\frac{(N-1)Q^0_{1ii}}{Q^0_{111}}\right)^\frac{3}{2}.
\label{eq:rsig}
\s]
A convenient way to parameterize $Q^0_{abc}$ is 
\s[
&Q^0_{111}=|Q^0| \cos \theta, \\
&Q^0_{1ii}=\frac{\left|Q^0\right|}{\sqrt{3(N-1)} }\sin \theta,\ (i=2,3,\ldots,N),
\label{eq:paraq}
\s]
where we can assume $0\leq \theta < \pi$ without loss of generality, because $\Psi(Q)=\Psi(-Q)$.  
Using \eq{eq:psigroup}, \eq{eq:rsig} and \eq{eq:paraq}, we obtain
\s[
\Psi\left(Q^0\right)\sim V \exp \left[ -2 I \left( \frac{k}{3} \right)^\frac{3}{2}  
\cos \theta \left( 1+\sqrt{\frac{N-1}{3}} \tan \theta \right)^\frac{3}{2}\left|Q^0\right| 
\right].
\label{eq:waveson}
\s]
Note that this formula is valid only within $0\leq \theta < \pi/2$, because 
$Q_{111}^0$ and $Q_{1ii}^0$ must have the same sign for the existence of the saddle point,
as commented below \eq{eq:qx}.

\begin{figure}
\begin{center}
\includegraphics[width=7cm]{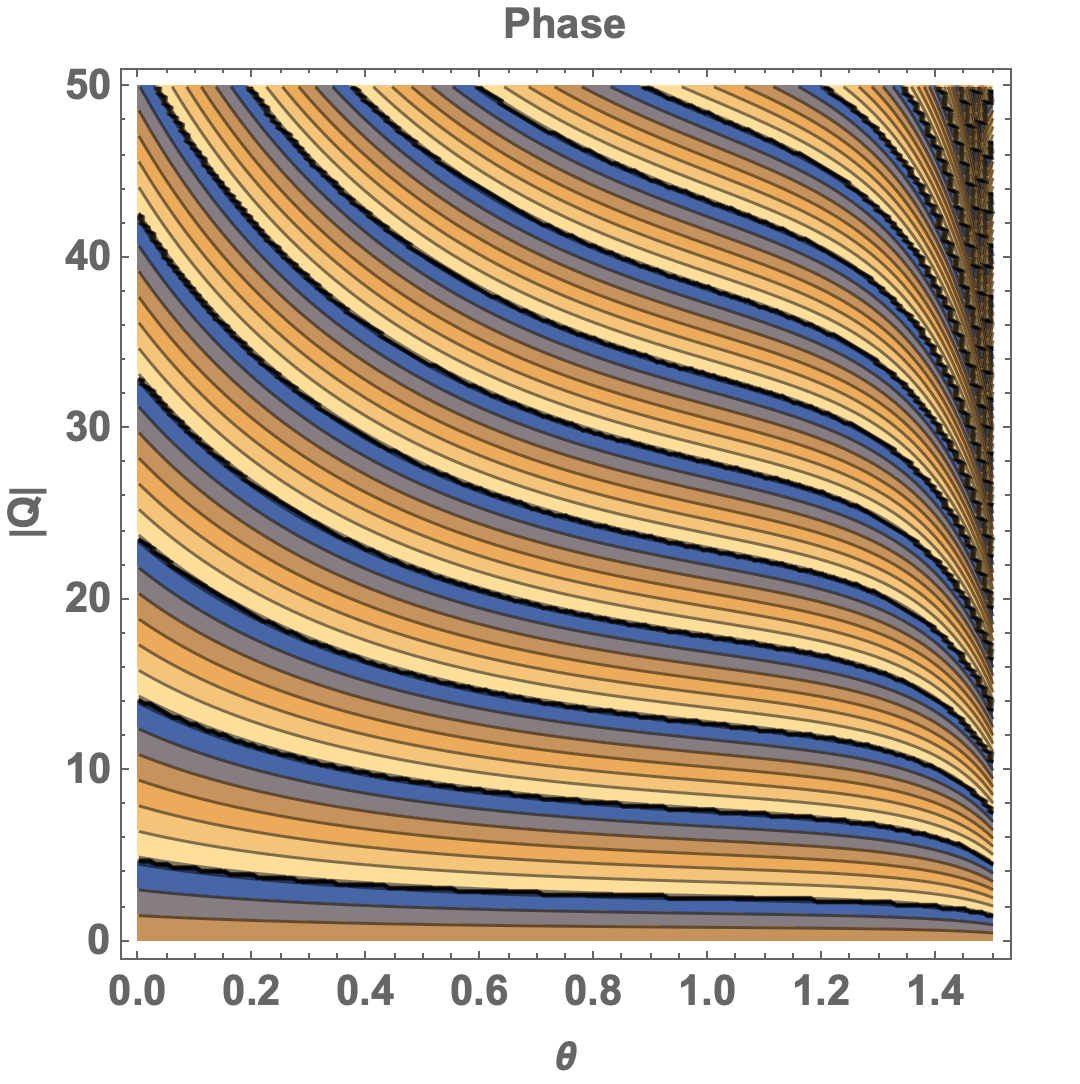}
\caption{The phase profile of \eq{eq:waveson} for $N=3$ and $k=3^{1/3}$
is shown by a contour plot of the phase modulo $2\pi$.   }
\label{fig:phaseN3ana}
\end{center}
\end{figure}

An example of the phase profile of \eq{eq:waveson} with $N=3$ is given in Figure~\ref{fig:phaseN3ana}, 
where $k=3^{1/3}$ is taken for later convenience.
If we assume that the evolutions are in the directions of increasing $|Q|$ and transverse to the constant
phase lines,  $\theta$ evolves toward $\pi/2$.
The question is how we can interpret the evolutions in physical terms.
In Section~\ref{sec:trd}, we will employ the tensor-geometry correspondence previously introduced 
through the tensor rank decomposition \cite{Kawano:2018pip}.

\section{Tensor rank decomposition and sizes of spaces}
\label{sec:trd}
Tensor rank decomposition \cite{Landsberg2012}
is to express a tensor by a sum of rank one tensors. More precisely, we 
express $Q_{abc}$ by
\[
Q_{abc}=\sum_{i=1}^{\tilde R} w_a^i w_b^i w_c^i, 
\label{eq:decomp}
\] 
where rank $\tilde R$ is taken to be the minimum value which realizes this decomposition. 
The tensor rank decomposition
is a tensor generalization of the singular value decomposition for a matrix. 
Note that, though \eq{eq:decomp} has the same form as the second equation in \eq{eq:saddleeq},
there is a crucial difference that the rank $\tilde R$ is the minimum value,
while $R$ in \eq{eq:saddleeq} can be any.

It is a difficult task to precisely obtain the decomposition \eq{eq:decomp}, though it is used for many applications.
This is different from the matrix case, in which the decomposition is unique (up to some trivial degeneracies) and has a well-posed definite procedure.
Hence a practical procedure of the tensor rank decomposition is to numerically optimize the vectors 
$w_a^i$ to approximately realize \eq{eq:decomp} assuming a value of $\tilde R$ 
as an input. If an optimized value of $w_a^i$ for an assumed value of $\tilde R$ 
is not good enough as an approximate realization of 
\eq{eq:decomp},  $\tilde R$ is increased and optimization is redone.
It is of course true that, if $\tilde R$ is large enough, one can always obtain an exact decomposition 
(up to a machine precision limit). 
However, in many cases, such a decomposition is too rough, namely,  there exist a lot of cancelations 
among the summands in \eq{eq:decomp}, and the decomposition does not well reflect
the true nature of $Q_{abc}$. 
Therefore, in many cases, it is better to take the smallest value of $\tilde R$ 
with an approximate decomposition up to an allowance one sets.

Let us now explain the tensor-geometry correspondence introduced in \cite{Kawano:2018pip},
namely, the procedure to interpret a given $Q_{abc}$ geometrically.
Let us assume a decomposition \eq{eq:decomp} is anyway obtained, approximately or not,
for a given $Q_{abc}$.
Differently from the matrix case, the vectors $w_a^i\ (i=1,2,\ldots,\tilde R)$ are not transverse among 
them in general. 
We suppose that each vector $w_a^i$ represents a ``point" $i$ forming a space, and inner products $w_a^i w_a^j$ 
represent neighborhood relations between points $i$ and $j$: As $w_a^i w_a^j$ is larger, the points $i$ and $j$
are nearer. By this procedure we obtain a geometric object
whose points have neighborhood relations.  
The validity of this procedure was checked for some simple but non-trivial examples 
in \cite{Kawano:2018pip}.

\begin{figure}
\begin{center}
\includegraphics[width=5cm]{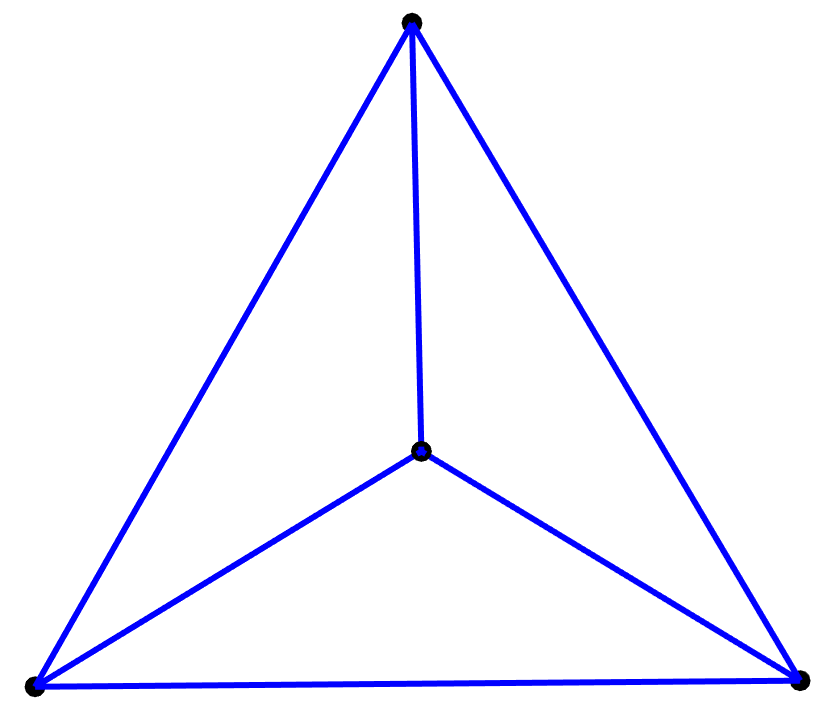}
\hfill
\includegraphics[width=5cm]{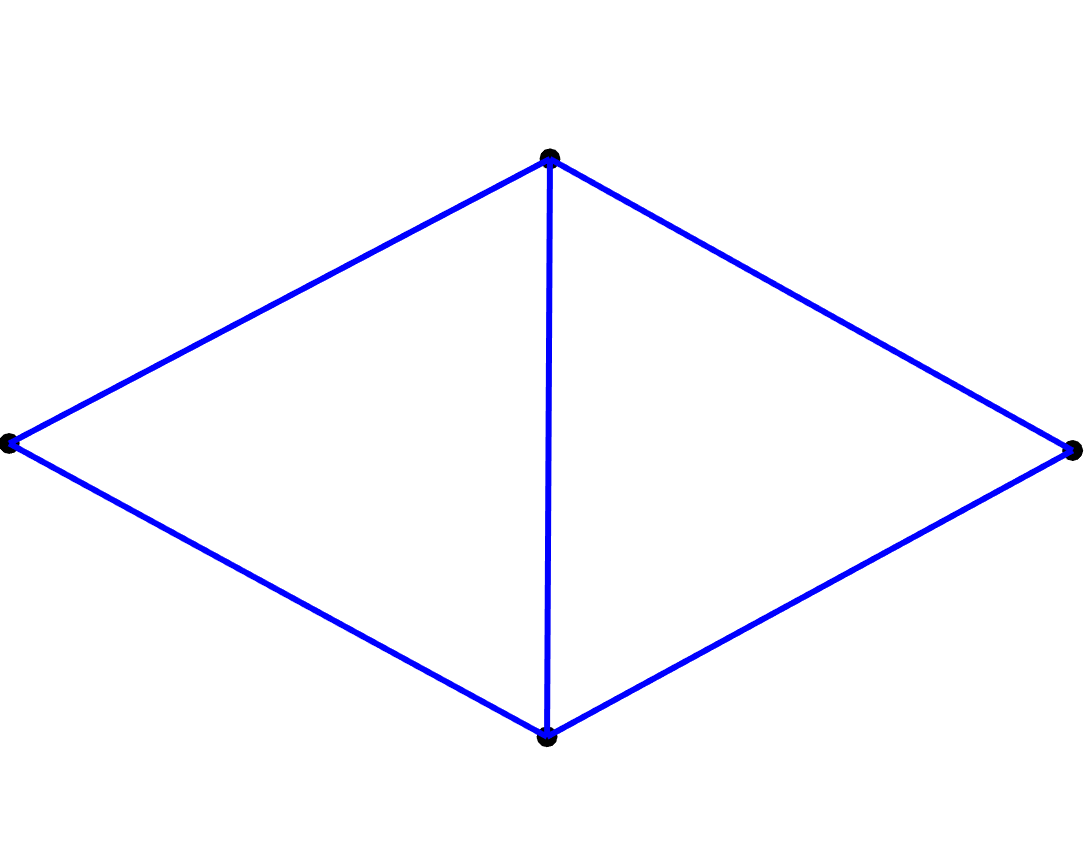}
\hfill
\includegraphics[width=5cm]{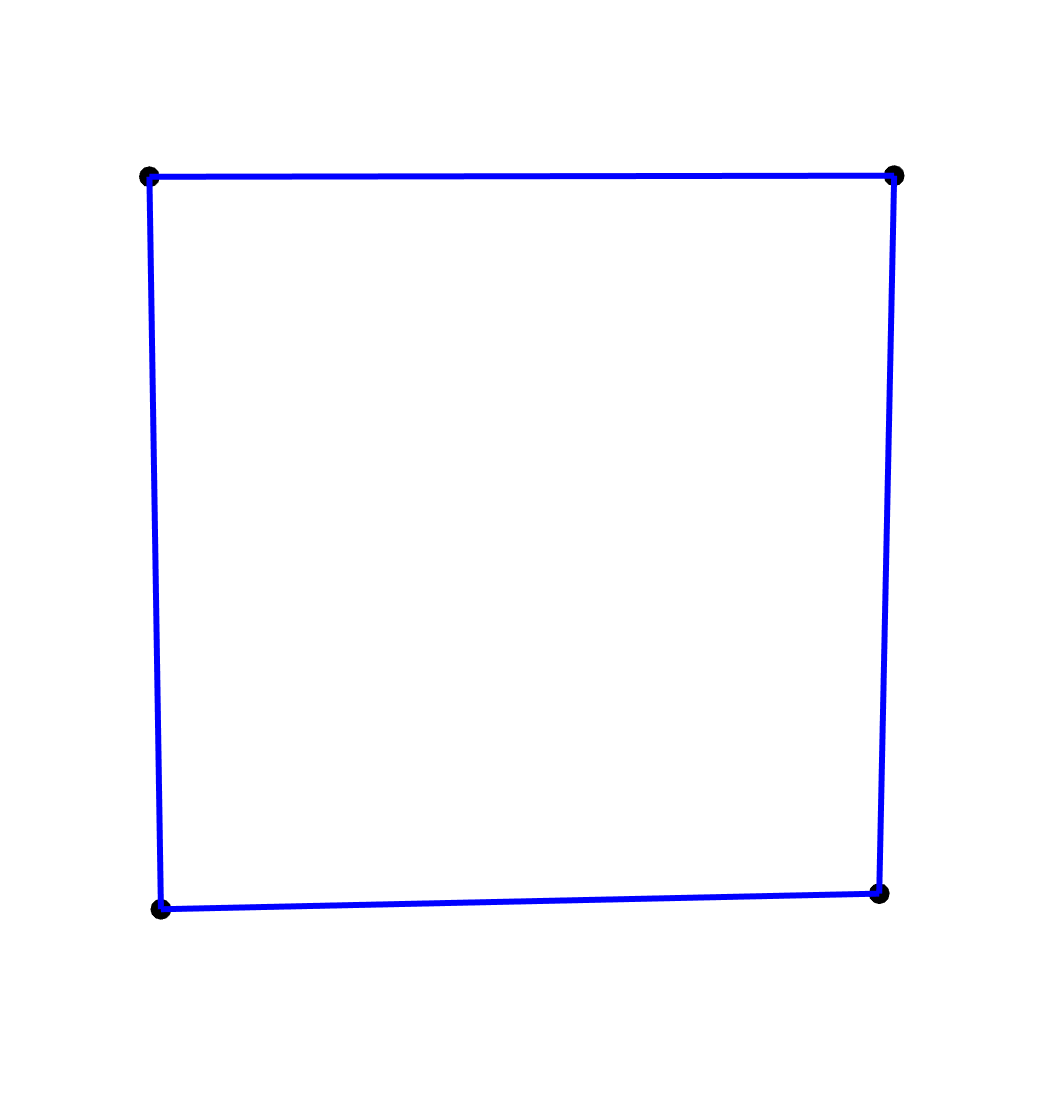}
\caption{The graphical representations of the tensor rank decomposition of the tensor \eq{eq:paraq}
for $N=3$.
Numerical optimizations show that $\tilde R=4$ is the minimal value which realizes \eq{eq:decomp} 
up to the machine 
precision limit. In the figures, there are four points because of $\tilde R=4$, 
and the edges are drawn between two points $i$ and $j$ with $w_a^i w_a^j>0$. 
The diagrams appear for 
$0.1 \lesssim \theta \lesssim 0.5$, $0.2 \lesssim \theta \lesssim 1.0$, and $1.1\lesssim \theta \lesssim 1.5$
from the left to the right, respectively.}
\label{fig:N3decomp}
\end{center}
\end{figure}

\begin{figure}
\begin{center}
\includegraphics[width=5cm]{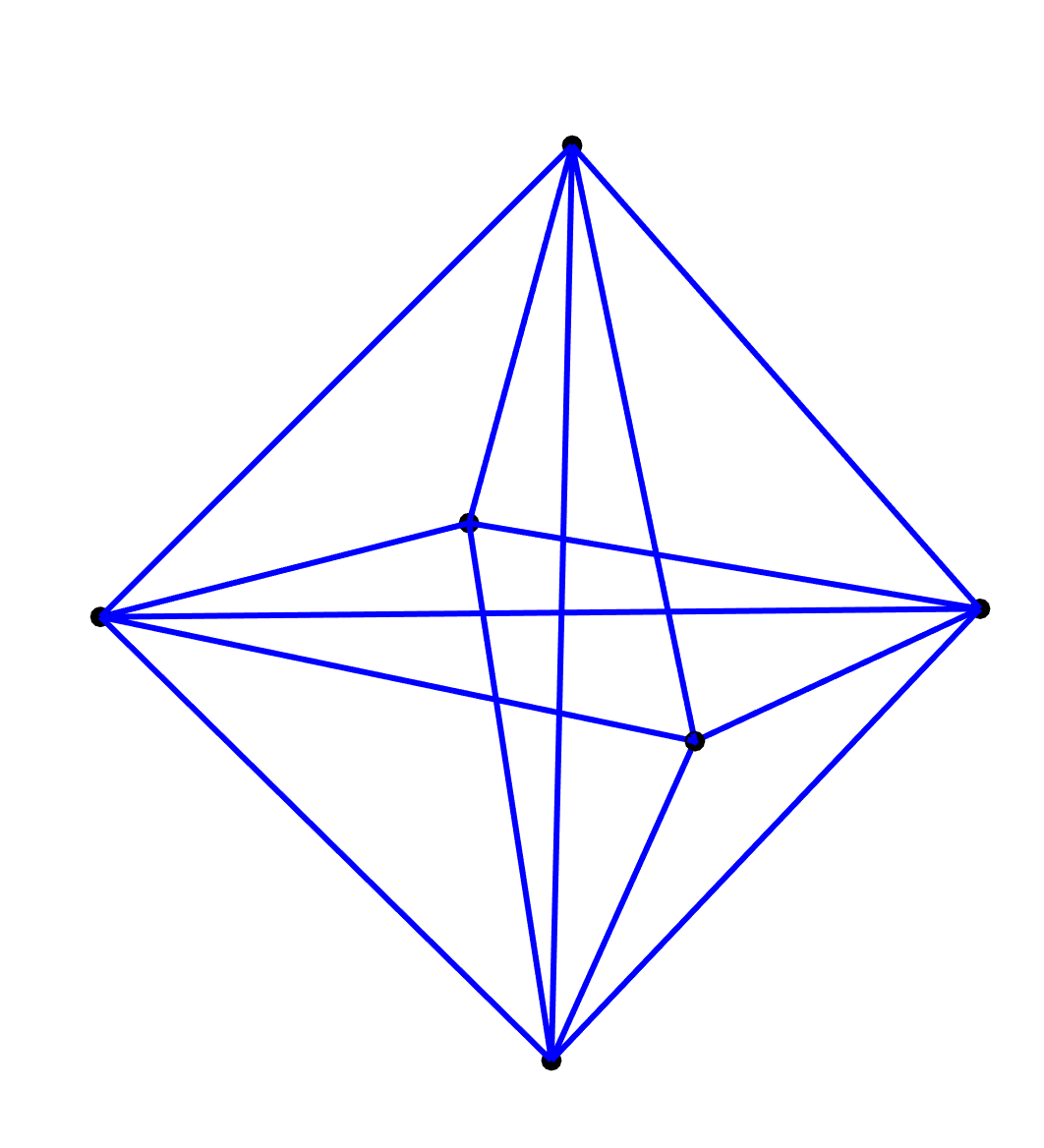} 
\hfill
\includegraphics[width=5cm]{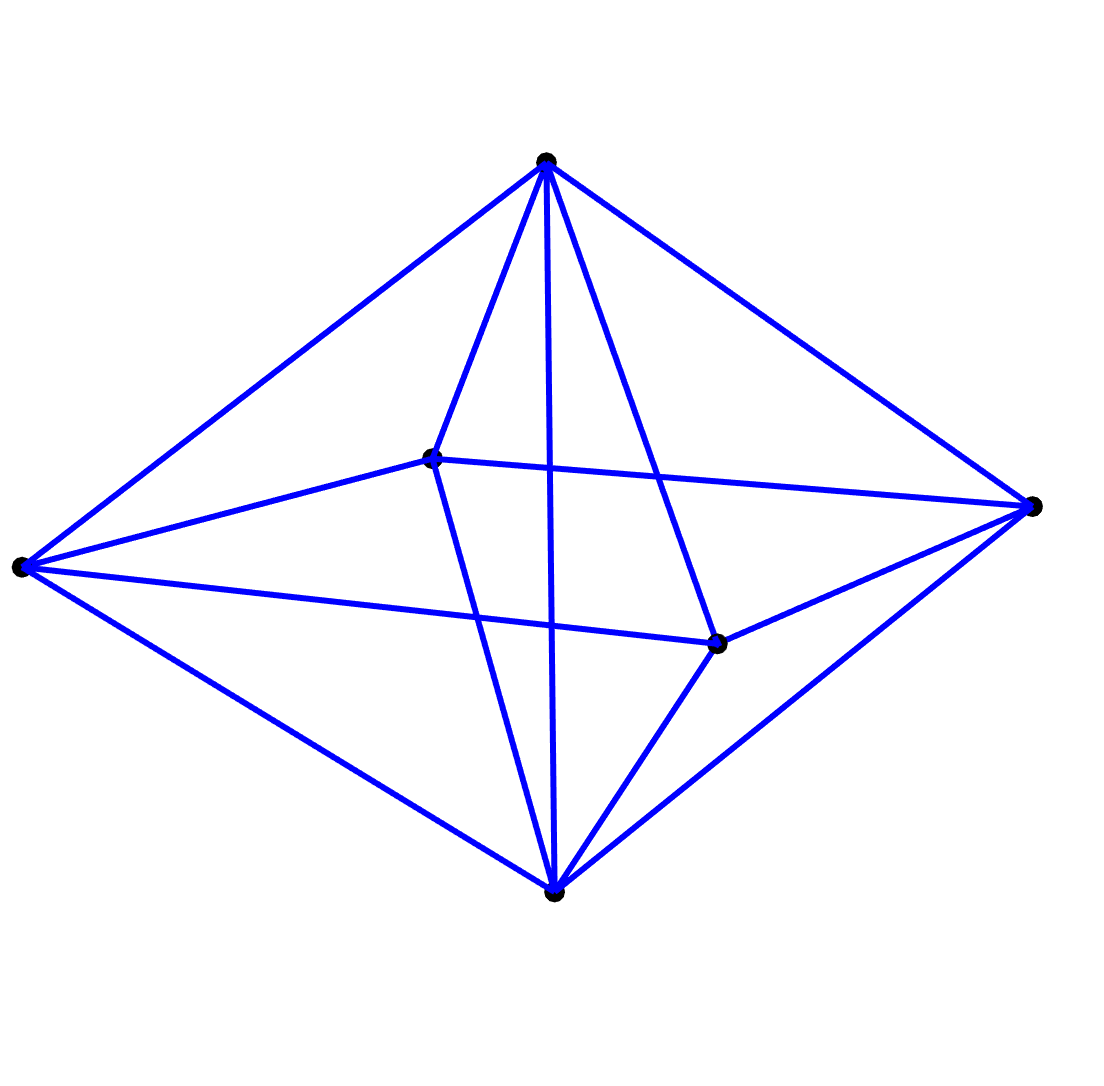}
\hfill
\includegraphics[width=5cm]{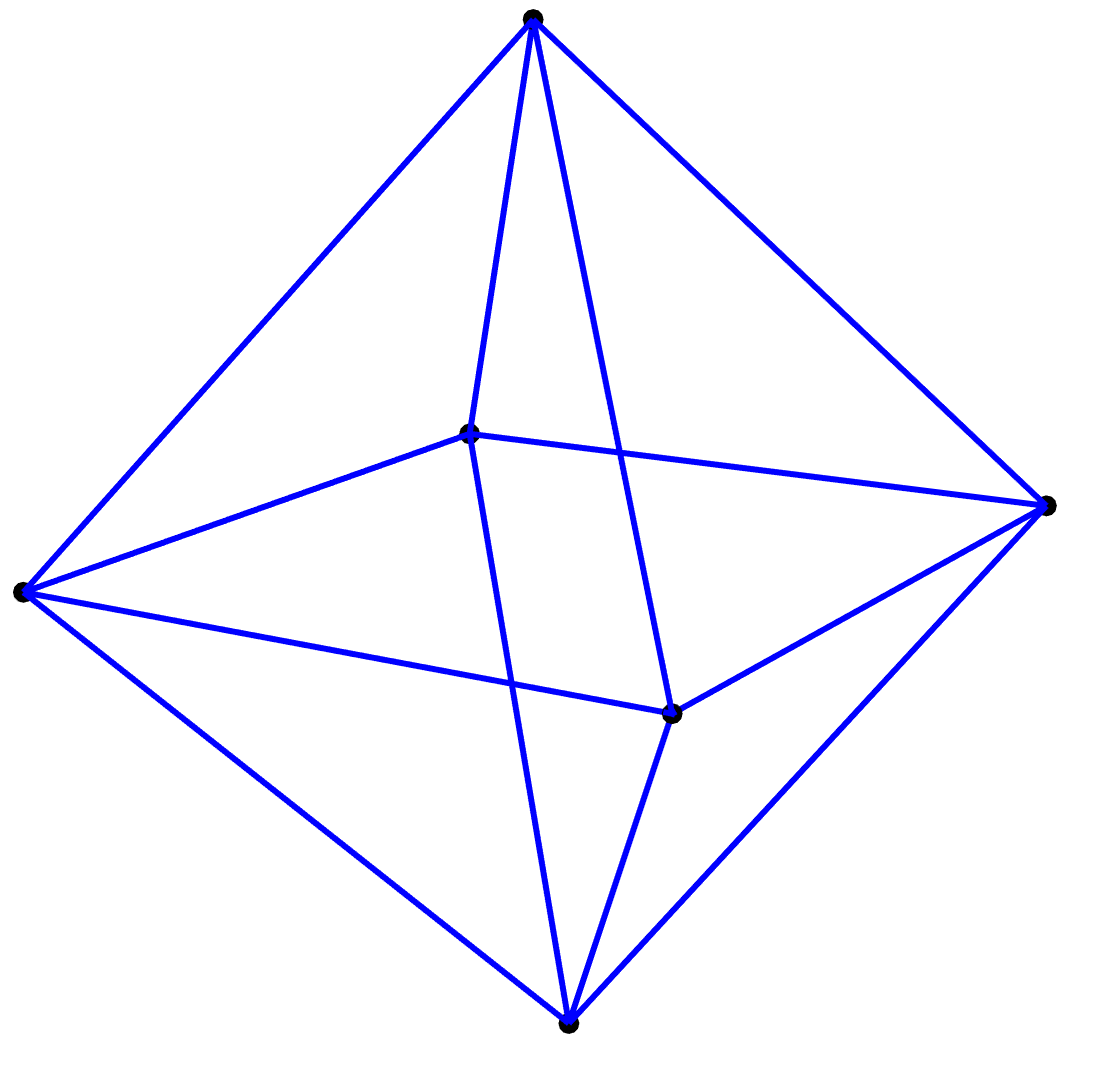}
\caption{The graphical representations of the tensor rank decomposition of the tensor \eq{eq:paraq}
for $N=4$, drawn in the same way as in Figure~\ref{fig:N3decomp}.
$\tilde R=6$ is the minimal value which realizes \eq{eq:decomp} up to the machine precision limit. 
The diagrams appear for $0.1 \lesssim \theta \lesssim 0.8$, $0.6 \lesssim \theta \lesssim 1.0$, and $1.1\lesssim \theta \lesssim 1.5$ from the left to the right, respectively.}
\label{fig:N4decomp}
\end{center}
\end{figure}

We apply the above procedure to find geometric objects corresponding to the tensor \eq{eq:paraq}.
The results are shown for $N=3$ in Figure~\ref{fig:N3decomp} and for $N=4$ in Figure~\ref{fig:N4decomp},
respectively. 
Unlike the general case explained above, the tensor rank decompositions for 
our examples can be clearly done, probably because of the smallness of $N$ and the high symmetry: 
The ranks can clearly be identified as $\tilde R=4$ for $N=3$ and $\tilde R=6$ for $N=4$, respectively.
The diagrams in the figures imply that the tensor \eq{eq:paraq} represents 
larger geometric objects as $\theta$ becomes larger within the range $[0,\pi/2]$: 
When $\theta$ is small, the points are mutiply connected, meaning that the objects are compact;
as $\theta$ is increased, inner edges disappear one by one, and eventually 
objects like $S^1$ for $N=3$ and $S^2$ for $N=4$ appear, respectively. 
The topologies of the final objects are consistent with the symmetry of the tensor, 
$SO(2)$ and $SO(3)$, respectively.  
An interesting observation is that, in both cases, the existing ranges of  the left two objects have 
overlapping regions of $\theta$, 
suggesting that there exist first-order transitions of geometries around the overlapping values 
of $\theta$. 

\section{Monte Carlo simulations for a few examples}
\label{sec:montecarlo}
In this section, we compute the phase profiles of 
the wave function for $N=3,4$ by Hamiltonian Monte Carlo simulations.
A motivation for this is to check the discussions in Section~\ref{sec:lie} and \ref{sec:profile}.  
Another is to see what really occurs beyond the approximations.  

Let us first specify the quantity we compute in the Monte Carlo simulations. The wave function \eq{eq:psiq}
is difficult to directly handle by Monte Carlo simulations, because it is complex. To make it more 
tractable we rewrite it into the form which can be computed by the re-weighting procedure.
By performing the $P_{abc}$ integral in \eq{eq:psiq}, we obtain delta functions, which we approximate  
by a Gaussian function. The integral over $\tilde \phi^i$ along ${\cal C}$ in \eq{eq:psiq} 
generates ${\rm Ai}(x)+I \,{\rm Bi}(x)$ with $x=-3^{-1/3}k\, \phi^i_a \phi^i_a$. 
Thus the quantity we compute in the simulations is the phase profile of the following function:
\[
\Psi_{MC}(Q)=\int_{{\mathbb R}^{NR}}
\prod_{a,i=1}^{N,R} d\phi_a^i\,
e^{ -\lambda \left(Q_{abc}- \sum_{i=1}^R \phi^i_a \phi^i_b \phi^i_c \right)
\left(Q_{abc}- \sum_{i=1}^R \phi^i_a \phi^i_b \phi^i_c \right) }
\prod_{j=1}^{R} \left\{{\rm Ai}(-\phi^j_a \phi^j_a)+I \, {\rm Bi}(-\phi^j_a \phi^j_a)\right\},
\label{eq:psimcq}
\]
where we have put $k=3^{1/3}$ for computational simplicity, 
and we simulate the delta functions by taking $\lambda$ large enough.

For the Monte Carlo simulations, it is more convenient to scale out the size of $Q_{abc}$. 
After parameterizing $Q_{abc}=|Q| \tilde Q_{abc}\ (|\tilde Q|=1)$ and performing 
a rescaling $\phi_a^i \rightarrow |Q|^{1/3} \phi_a^i$, we obtain
\s[
\Psi_{MC}(Q)&=|Q|^{NR/3}\int_{{\mathbb R}^{NR}}
\prod_{a,i=1}^{N,R} d\phi_a^i\,
e^{ -\tilde \lambda  \left(\tilde Q_{abc}- \sum_{i=1}^R \phi^i_a \phi^i_b \phi^i_c \right)
\left(\tilde Q_{abc}- \sum_{i=1}^R \phi^i_a \phi^i_b \phi^i_c \right) } \\
&\hspace{3cm}
\cdot \prod_{j=1}^{R} \left\{{\rm Ai}\left(-|Q|^{2/3}\phi^j_a \phi^j_a\right)
+I \, {\rm Bi}\left(-|Q|^{2/3}\phi^j_a \phi^j_a\right)\right\}.
\label{eq:psiqres}
\s]
where $\tilde \lambda:=\lambda |Q|^2$.
Then the re-weighting procedure of Monte Carlo simulation is to
compute the average 
\[
\left \langle \prod_{i=1}^{R} \left\{{\rm Ai}\left(-|Q|^{2/3}\phi^i_a \phi^i_a\right)
+I \, {\rm Bi}\left(-|Q|^{2/3}\phi^i_a \phi^i_a\right)\right\}
\right \rangle,
\label{eq:average}
\]
where the sampling of $\phi_a^i$ is performed by the weight, 
\[
e^{ -\tilde \lambda  \left(\tilde Q_{abc}- \sum_{i=1}^R \phi^i_a \phi^i_b \phi^i_c \right)
\left(\tilde Q_{abc}- \sum_{i=1}^R \phi^i_a \phi^i_b \phi^i_c \right) },
\label{eq:weight}
\]
for a given $\tilde Q_{abc}$. 
Since we are only interested in the phase profile, we only compute \eq{eq:average},
and do not compute the overall weight, namely, the integration of \eq{eq:weight} over $\phi_a^i$,
which is real.

In our simulations, we take $\tilde \lambda$ as an input parameter rather than $\lambda$.
One reason is that $\lambda$ was introduced to approximate the delta functions 
by taking large $\lambda$, and therefore it is a sort of virtual parameter whose value itself
is not important. The other reason, which is very practical, is that we can compute  
the phase profile over various values of $|Q|$ by computing \eq{eq:average} from
one set of sampling with the weight \eq{eq:weight} with a $\tilde \lambda$. 
This saves much runtime:
If one took $\lambda$ as an input parameter, then $\tilde \lambda$ would depend on $|Q|$
and we would have to take sampling for every value of $|Q|$. 
It is expected that the phase profile does not depend much on which of $\lambda$ or 
$\tilde \lambda$ is chosen as an input parameter, 
if it is taken large enough.\footnote{In fact, this can be observed in the actual
Monte Carlo results.} 
Note that the overall factor may depend on the choice, but the dependence will be
a real factor irrelevant for the phase profile.

\begin{figure}
\begin{center}
\includegraphics[width=7cm]{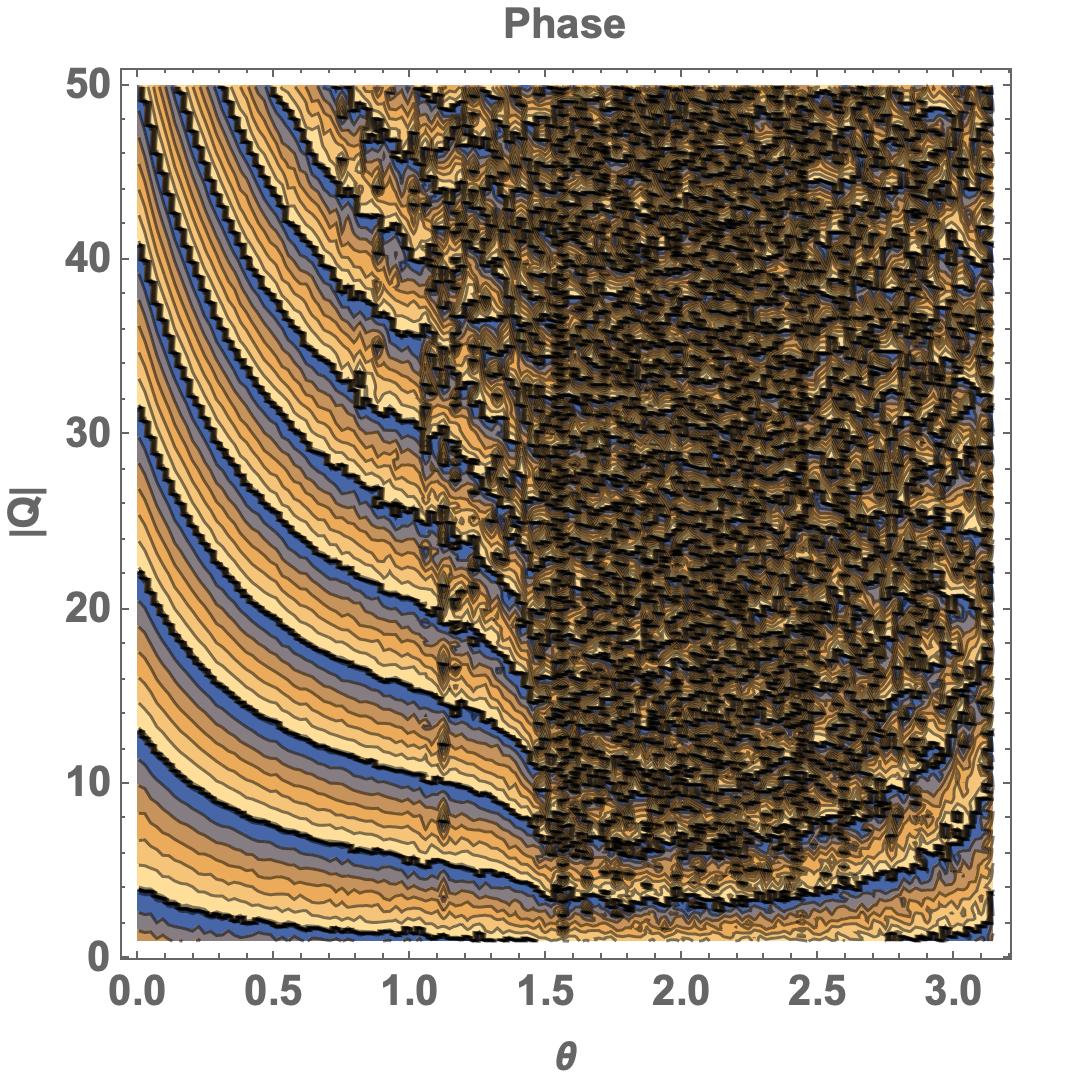}
\hfil
\includegraphics[width=7cm]{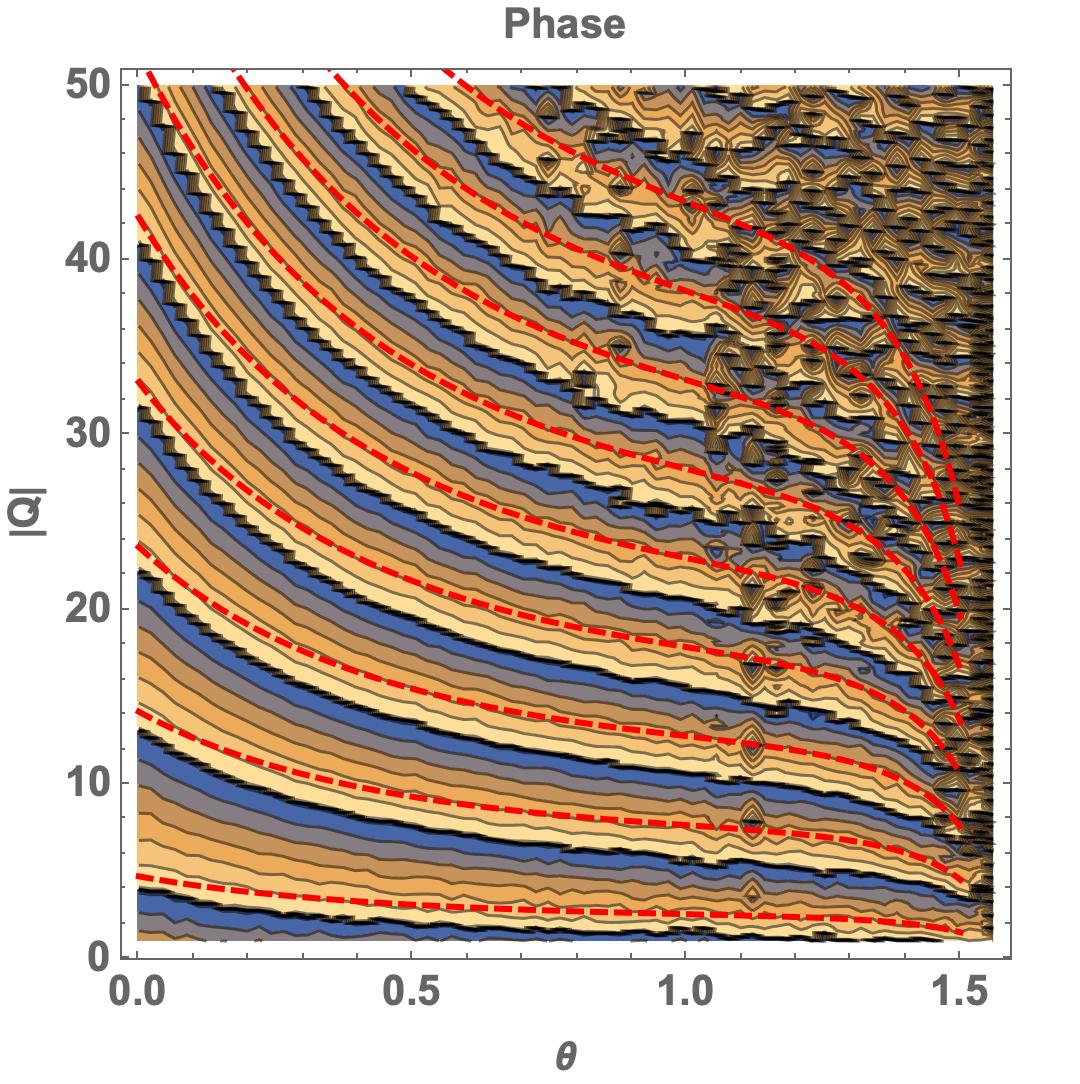}
\caption{The phase profile for \eq{eq:paraq} with $N=3$, $R=7$,  
$k=3^{1/3}$ and $\tilde \lambda=10^7$, 
obtained by computing \eq{eq:average} by the Hamiltonian Monte Carlo method.
The number of $\phi_a^i$ samples is $5\cdot 10^4$ for each $\tilde Q_{abc}$.
The pictures are drawn in the same manner as in Figure~\ref{fig:phaseN3ana}. 
The right figure shows the half range $0\leq \theta < \pi/2$ of the left one. 
The dashed lines are the constant phase lines ($\pi$ modulo $2 \pi$) 
derived from \eq{eq:waveson}. }
\label{fig:N3SO2}
\end{center}
\end{figure}

Figure~\ref{fig:N3SO2} shows the phase profile for \eq{eq:paraq} with $N=3$, which corresponds to
Figure~\ref{fig:phaseN3ana} derived from the saddle point analysis.  Here we take $R=7$,
which is one of the nearest values to the real one $(N+2)(N+3)/4=7.5$.
The phase profile looks regular only in the region $0\leq \theta <\pi/2$, 
which is the only region with the saddle point solution, as commented below \eq{eq:waveson}. 
In the other region $\pi/2\leq \theta<\pi$, 
the result of phase profile seems to have no sensible structure, looking like random.
As can be seen in the right figure, the constant phase lines derived from the saddle point analysis 
\eq{eq:waveson} agree well with the simulation result.  However, 
the phase profile is disrupted in the region near $\theta \sim \pi/2$ with large $|Q|$. 
As discussed in Section~\ref{sec:trd}, this region corresponds to the large spacetime
region. 
We are tempted to suspect that there exists some light modes generating fluctuations
in the region.

In Figure~\ref{fig:N3SO2} one can notice the tendency that disruptions of the phase profile are 
vertically aligned. This is not a physical effect, but is just because we use the 
same sampling data of $\phi_a^i$ to compute \eq{eq:average} for different values of $|Q|$ 
and a fixed value of $\theta$. If a sampling data happens to contain 
some irregularities (of statistical origins), they affect all the phase profiles vertically aligned. 

\begin{figure}
\begin{center}
\includegraphics[width=7cm]{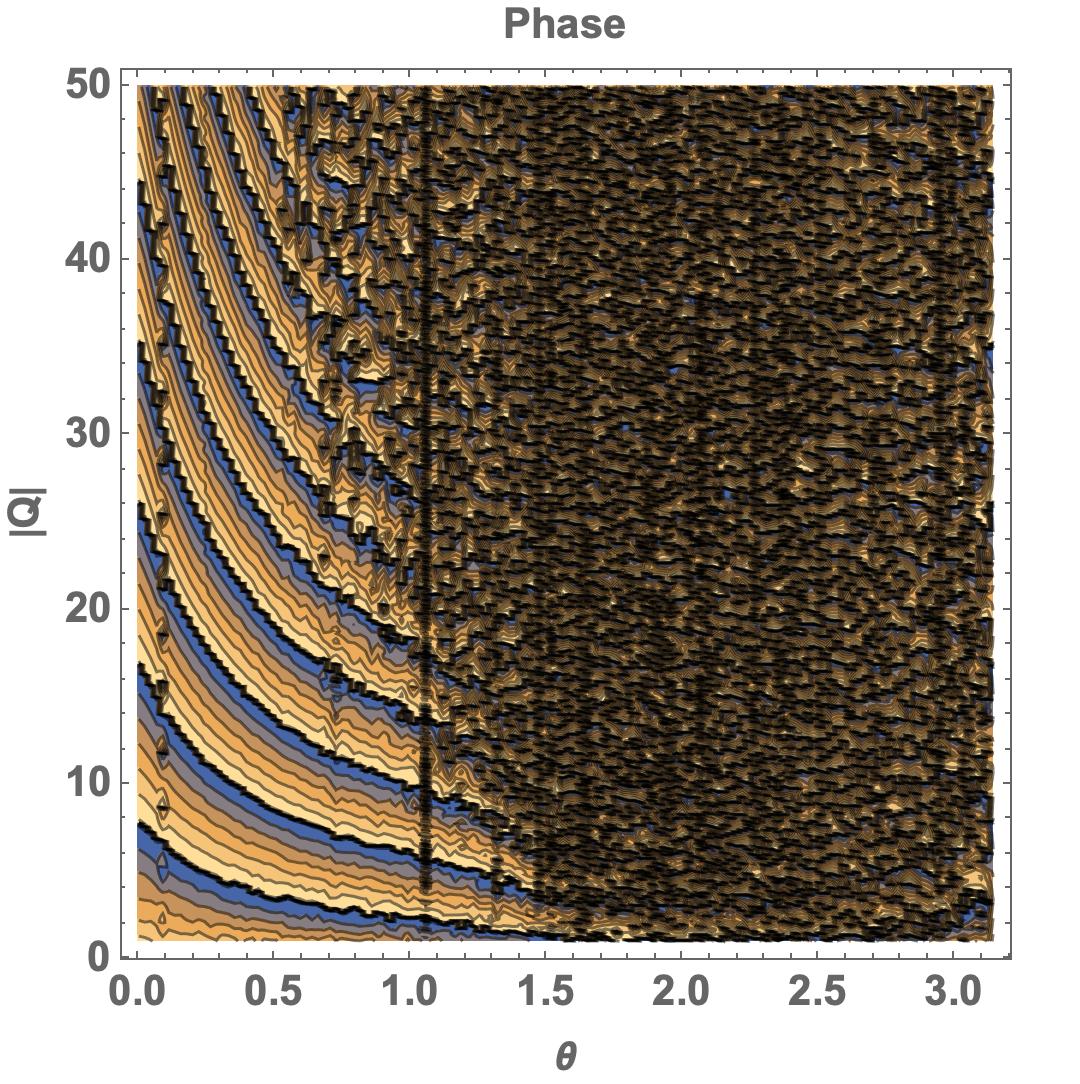}
\hfil
\includegraphics[width=7cm]{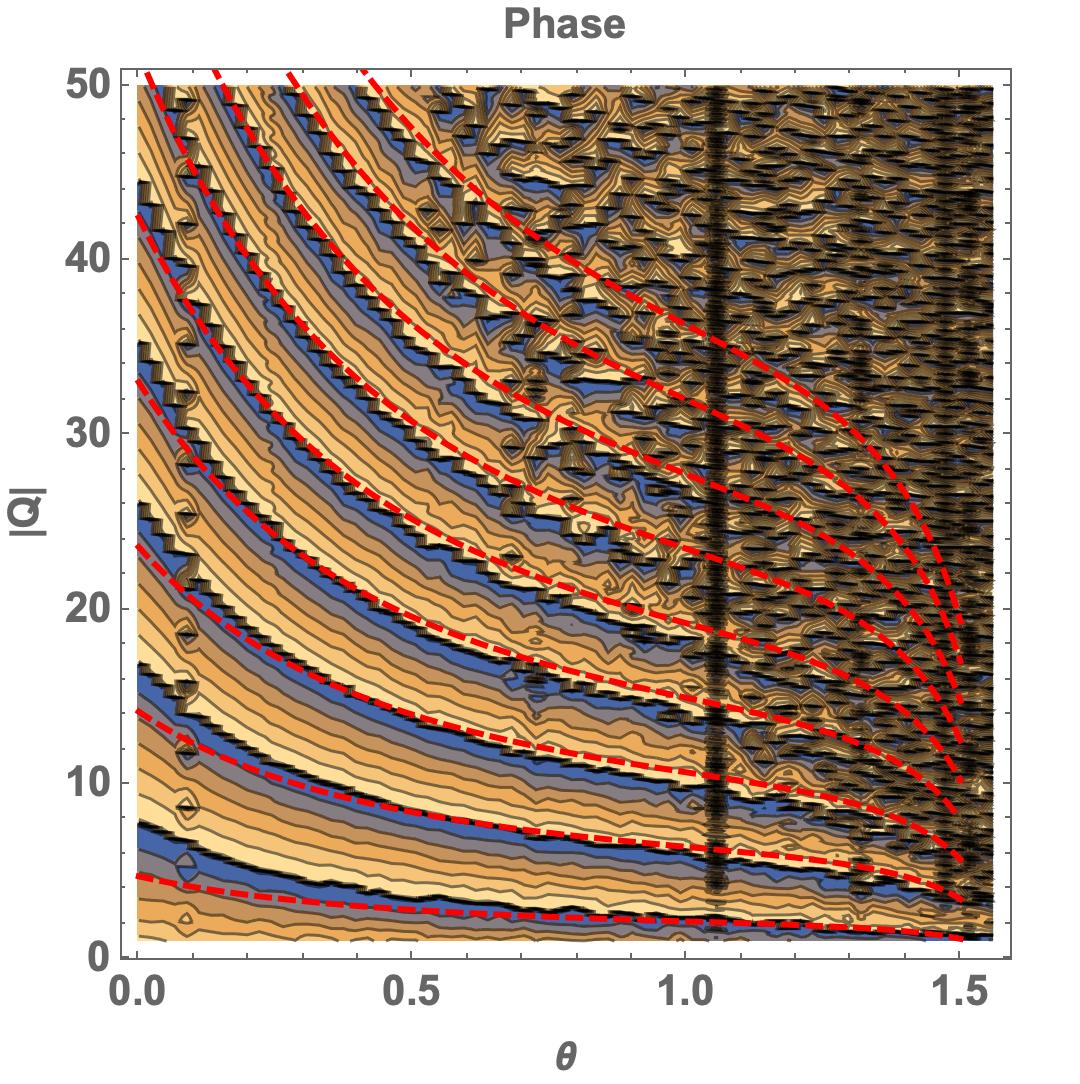}
\caption{The phase profile for \eq{eq:paraq} with $N=4$, $R=10$,  
$k=3^{1/3}$, $\tilde \lambda=10^7$, and the sample number is $10^5$ for each $\tilde Q_{abc}$.
The pictures are drawn in the same manner as the $N=3$ case in Figure~\ref{fig:N3SO2}. }
\label{fig:N4SO3}
\end{center}
\end{figure}

We show the phase profile for the $N=4$ case in Figure~\ref{fig:N4SO3}.
We needed to take a larger number of sampling than 
the $N=3$ case in order to obtain a phase profile which is more or less as regular as the $N=3$ 
case.
 
\begin{figure}
\begin{center}
\includegraphics[width=7cm]{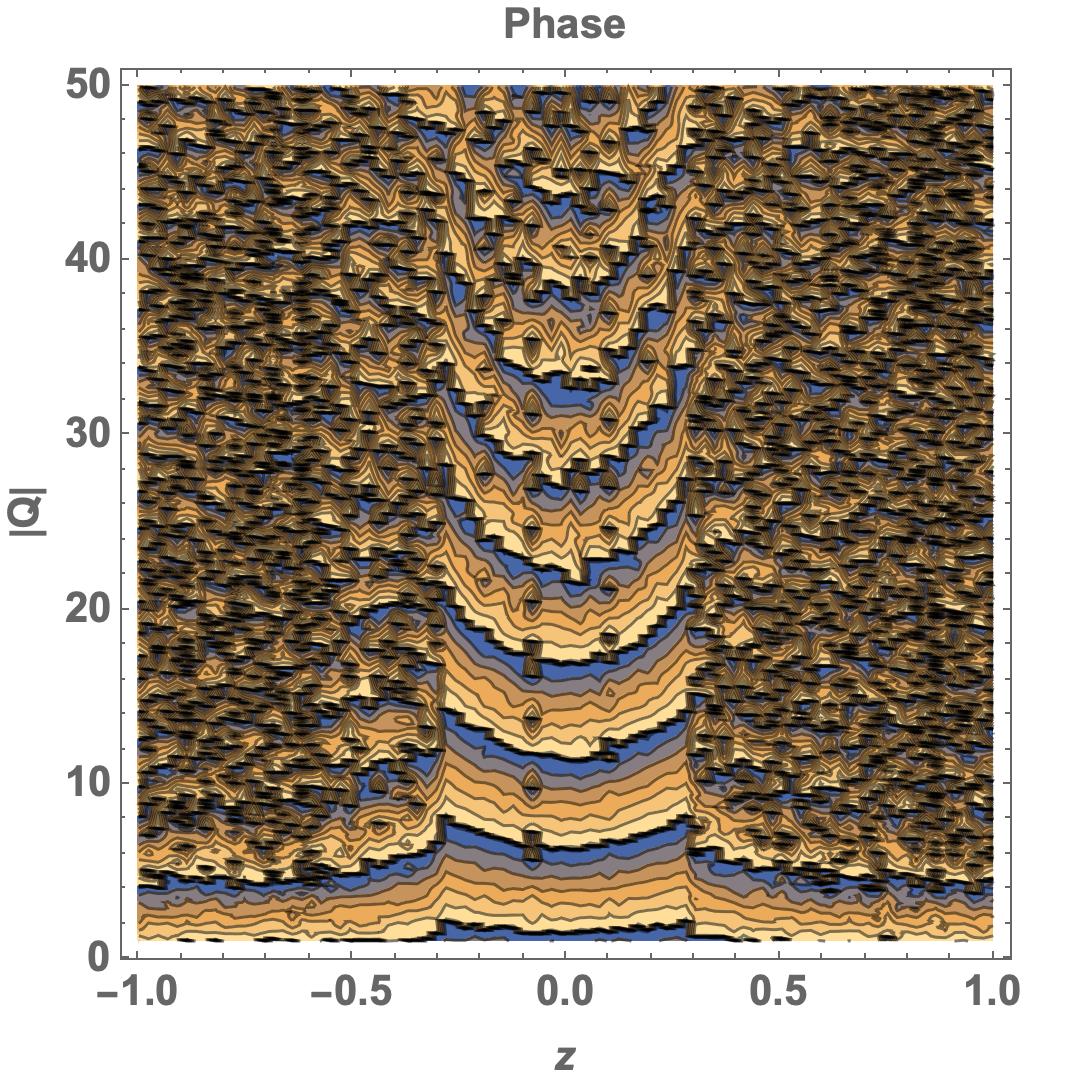}
\caption{The phase profile for a symmetry breaking example.
The horizontal axis is the breaking parameter $z$ in \eq{eq:paraqz}.
The other parameters are $\theta=\pi/4$, $R=7$,  
$k=3^{1/3}$ and $\tilde \lambda=10^6$. The number of $\phi_a^i$ sampling is $10^4$ 
for each $\tilde Q_{abc}$. 
}
\label{fig:N3z}
\end{center}
\end{figure}

Figure~\ref{fig:N3z} shows the phase profile for tensors which break the Lie group symmetry. More
precisely, we consider $N=3$ and introduce a breaking parameter $z$ to deform \eq{eq:paraq}:
\s[
&Q_{111}=q \cos \theta, \\
&Q_{122}=q \left( \frac{1}{\sqrt{6} }\sin \theta-z\right),\\
&Q_{133}=q \left( \frac{1}{\sqrt{6} }\sin \theta+z\right).
\label{eq:paraqz}
\s] 
One can find that the regularity of the phase profile appears only around the symmetric tensor values
$z=0$.  The constant phase lines of the phase profile also suggest that the direction of evolution 
is toward $z=0$. 

\section{Summary and outlook}
As explained in Section~\ref{sec:introduction}, 
it is important to study the wave function in the coordinate ($Q_{abc}$) representation
to understand the spacetime dynamics in the canonical tensor model (CTM).
This paper is the first such study, while the previous analyses have been done in the 
momentum ($P_{abc}$) representation. 
We have computed the phase profile of the wave function in an analytic method with saddle point
analysis and a numerical method with Hamiltonian Monte Carlo simulations.
In fact, the phase profile is an appropriate quantity for the first study,
since this is the easiest to compute in both methods.
We have found that Lie group symmetric spacetimes are strongly favored, and 
the phase profile suggests that sizes of spacetimes grow in ``time".  
We have also observed that the phase profile is disrupted in the large spacetime region,
that would suggest existence of light modes generating fluctuations in the region.

The mechanism which makes Lie group symmetric configurations dominant 
in the coordinate representation is more interesting than the similar phenomenon 
in the momentum representation. We have found that, 
for Lie group symmetric $Q_{abc}$, continuum description naturally appears
giving a fluid picture. Pursuing an effective theory of the fluid in future study would be a feasible 
strategy toward understanding the spacetime dynamics of CTM for large $N$, 
considering that it would be too complicated to study CTM in the genuine discrete form.
In fact, there have been discussions mapping gravity to fluid dynamics in the context 
of analogue gravity \cite{Barcelo:2005fc}.
It would be highly interesting, if there are connections between the fluid dynamics of CTM
and general relativity. 

One of the subtle points in this paper is that we assumed that the evolution of CTM 
occurs in the direction transverse to the constant phase surfaces of the wave function.
This assumption was used to argue that sizes of spaces grow in ``time". 
This is a commonly accepted assumption in ordinary systems, but we do not know whether 
this is valid for CTM. 
To be more definite, we have to introduce ``time" to CTM in some manner 
and discuss spacetime evolutions with this ``time" variable.   
Since the formalism of CTM is analogous to the ADM formalism of general relativity, 
we would be able to apply to CTM the proposals made previously 
in the context of quantized general relativity in the literature \cite{Isham:1992ms}.  

Another question to be answered is to clarify the reason for the disruptions of the phase
profile observed in the large spacetime region.
In fact, we have performed other simulations than shown explicitly in the paper, but 
it was hard to considerably suppress the disruptions by just performing simulations 
with larger $\tilde \lambda$ and more sampling.
It is plausible that the disruptions are related to an intrinsic character of CTM,
such as presence of light modes, but currently we have not totally 
excluded the possibility that it just comes from the bad performance of the 
re-weighting procedure containing a lot of cancelations.    
We hope that understanding of the fluid dynamics mentioned above would help us to 
get some insights to this problem.   

If there exist such light modes generating fluctuations, it would be interesting to try to compute 
primordial fluctuations from the perspective of CTM. Note that the beginning of the universe
from the perspective of CTM seems to be different from the standard inflation scenario with some
scalar fields. In CTM, the shape $Q_{abc}/|Q|$ can take any values with equal weights initially at 
$|Q|=+0$. Then as $|Q|$ becomes larger, only Lie group symmetric $Q_{abc}/|Q|$ are quantum
mechanically selected, and the size of a space becomes larger. 
Therefore the beginning of the universe in CTM is a sort of discarding process of non-symmetric 
configurations, while it is a sort of dominating process of a uniform symmetric configuration 
in the inflation scenario. Computation of primordial fluctuations and comparison with 
astrophysical datas would provide an interesting means to discriminate the two scenarios.  

\vspace{.3cm}
\section*{Acknowledgements}
The work of N.S. is supported in part by JSPS KAKENHI Grant No.19K03825. 


\end{document}